\documentclass[11pt]{article}

\usepackage{fullpage}
\usepackage{graphicx}
\graphicspath{{figs/}}
\usepackage{subcaption}
\usepackage{amsmath,amssymb,amsfonts, amsthm}
\usepackage{xcolor}
\usepackage{xspace}
\usepackage{mathtools}
\usepackage[store-sets-label]{keytheorems}

\usepackage{xcolor}
\usepackage{enumerate}
\usepackage{enumitem}
\usepackage{xspace}
\usepackage[T1]{fontenc}
\usepackage{subfiles}
\usepackage{bm}
\usepackage{tabularx}
\usepackage{multirow}
\usepackage{hyperref}
\usepackage{lineno}
\usepackage{titling}

\usepackage[symbol]{footmisc}

\hypersetup{
    colorlinks,
    linkcolor={red!50!black},
    citecolor={blue!50!black},
    urlcolor={blue!80!black}
}
\usepackage[capitalize,noabbrev]{cleveref}
\usepackage{soul}

\usepackage{algorithm}
\usepackage{algpseudocode}
\algtext*{EndWhile}
\algtext*{EndIf}
\algtext*{EndFor}
\algrenewcommand\algorithmicrequire{\textbf{Input:}}
\algrenewcommand\algorithmicensure{\textbf{Output:}}

\usepackage[colorinlistoftodos,textsize=tiny,textwidth=2cm,color=red!25!white,obeyFinal]{todonotes}

\usepackage{tikz}
\usetikzlibrary{fit,shadows,positioning,shapes.misc,decorations.pathreplacing,calc}

\usepackage{enumitem}
\setlist{topsep=3pt, itemsep=0pt}

\newkeytheorem{theorem}
\newkeytheorem{lemma,definition,claim,observation,corollary,remark}[sibling=theorem]

\newcommand{\ceil}[1]{{\left\lceil#1\right\rceil}}

\allowdisplaybreaks

\usepackage{graphicx} %
\usepackage{soul}
\title{Deterministic Distributed Graph Connectivity Sketching: \\Beating the Naive Sketch}

\thanksmarkseries{alph}

\author{ 
{Jacob Holm\thanks{University of Copenhagen, Email: \texttt{jaho@di.ku.dk}.}}
\and
{David Rasmussen Lolck\thanks{Institute of Science and Technology Austria, Email: \texttt{david.lolck@ista.ac.at}.
A part of this work was done during affiliation with University of Copenhagen.}}
}
\date{\vspace{-9.5ex}}

\begin{document}

\maketitle

\begin{abstract}
In the distributed graph sketching model, each vertex in the graph is considered as a separate processor. Each vertex is given access to its neighbourhood, and is then responsible for composing a message to a central referee, with the goal being minimising the length of these messages. Classic problems in this model include computing spanning forests, connected components and connectivity of the full graph. 

In the case where the vertices have to act deterministically, it has long been suspected that the correct upper-bound for computing spanning forests, connected components or connectivity would be the naive method of reporting every edge, which can be done with per-vertex message length of \(\lfloor n/2\rfloor\) bits. In this paper we show that this is not true for connected components and connectivity, by presenting a protocol that only requires \(n/3 +3\log_{2}n\) bits per vertex for \(n\geq13\).

We also show that in this setting connected components and connectivity are closely related in the required message length of the vertices. Letting \(C_{cc}(n)\) be the minimum required message length in bits per vertex to compute the connected components in a graph with \(n\) vertices and letting \(C_{conn}(n)\) be the minimum message length to compute if the graph is connected, we show that \(C_{cc}(n) \le C_{conn}(2n)\) with a multiplicative blow-up in the error probability of \((n-1)^2\).
\end{abstract}

\clearpage
\section{Introduction}
Graph sketching asks how much information about a graph must be retained in order to recover a prescribed global property.  We study the \emph{distributed graph-sketching model}, closely related to the linear graph-sketching framework of Ahn, Guha, and McGregor~\cite{agm12linear,agm12sketch}.  The input is a simple undirected graph \(G=(V,E)\).  Each vertex \(v\) is viewed as a separate player that knows only its own neighbourhood \(N_G(v)\).  Without communicating with the other vertices, every player sends a single message to a central referee, and the referee must compute the desired property of \(G\) from the labelled collection of messages.  The cost of a protocol is the maximum message length, measured in bits, over all vertices and all possible local inputs.  Equivalently, this is a simultaneous-message protocol specialised to graph inputs; it is also closely connected to one-round algorithms in the broadcast congested clique~\cite{yu2021connectivity,robinson2023distributed}.

Our focus is connectivity.  More precisely, we ask the referee to recover the partition of \(V\) into the connected components of \(G\).  This task is stronger than merely deciding whether \(G\) is connected, but weaker than outputting a spanning forest.  It is a natural test case for local sketching: connectivity is determined by the interaction of edges throughout the graph, while no individual vertex likely has enough information to certify the global answer.

The amount of randomness available to the players has a decisive effect on the known bounds.  With shared public randomness, the AGM framework gives sketches of \(O(\log^3 n)\) bits per vertex from which the referee can recover a spanning forest with high probability~\cite{agm12linear}.  With only private randomness, Holm, King, Thorup, Zamir, and Zwick obtain an \(O(\sqrt n\log n)\)-bit protocol~\cite{hktzz19}.  We instead study exact deterministic protocols, in which all encoding maps and the referee are fixed in advance.

A simple deterministic protocol uses \(\lfloor n/2\rfloor\) bits per vertex.  Fix a near-regular orientation of the complete graph on \(V\), known to every player and to the referee, and assign each potential edge to its tail.  Vertex \(v\) reports one bit for every potential edge assigned to it, indicating whether that edge is present in \(G\).  Since the orientation has maximum out-degree \(\lfloor n/2\rfloor\), the referee reconstructs the entire graph within this message bound. In~\cite{hktzz19} the authors observed that this is the best known deterministic bound for producing a spanning forest and asked whether it is optimal.

Our main result shows that the \(n/2\) barrier is not inherent for connectivity or connected-component recovery.  For \(n\ge 13\), we give an exact deterministic sketch using at most
\(\frac{n}{3}+3\lg n\)
bits per vertex.  The construction decomposes the complete graph into carefully balanced copies of \(K_{3,k}\).  Each such bipartite piece admits a component sketch that is shorter than its raw incidence vector, and the local sketches over all pieces can be concatenated without exceeding the stated budget.  The referee first recovers the connected components inside every piece and then takes the transitive closure of these local component relations to obtain the components of the full graph.

This result does not settle the deterministic spanning-forest question, since recovering a component partition need not identify any actual forest edges.  It does, however, separate that question from the weaker connectivity tasks and improves the deterministic upper bound for both connectivity and connected-component recovery by a constant factor.  We also show that deciding connectivity and recovering all connected components are equivalent up to replacing an \(n\)-vertex instance by a \(2n\)-vertex instance and worsening the error probability.  Thus, at the level of communication complexity, the decision and component-recovery versions are much closer than their output requirements might initially suggest.

\subsection{Our Results and Technical Overview}
We begin by isolating a simple principle that lets us reconstruct global connectivity from connectivity information on a family of smaller subgraphs.  Let \(K_V\) denote the complete graph on a vertex set \(V\), and let \(\omega:E(K_V)\to\{0,1\}\) specify which potential edges are present.  For \(F\subseteq E(K_V)\), write
\[
    F^{\omega=1}=\{e\in F: \omega(e)=1\}.
\]
\begin{lemma}[store=lem:uncover]
    Given a complete graph \(G=(V,E,\omega)\) and a covering of the edges \(E=E_1\cup\cdots\cup E_n\), then given the connected components of \(G_i=(V,E_i^{\omega=1})\) for each \(1\le i \le n\) there exists an algorithm computing the connected components of \(G=(V,E^{\omega=1})\).
\end{lemma}
\begin{proof}
    Follows from transitivity of connectivity.
\end{proof}
From this, the general goal is to find some decomposition into graphs that are easier to handle than the complete graph of \(n\) vertices. Our choice of graph is the complete bipartite graph \(K_{3,t}\) of \(3\) vertices on the left side and \(t\) vertices on the right.

\begin{lemma}[store=lem:algorithm]
    Let \((L\uplus R,E, \omega)\) be a complete bipartite graph where \(\omega\) labels each of the edges either \(0\) or \(1\), with \(|L|=3\) and \(|R|=k\). Then there exist collections of maps \(m_v^L : 2^R\to [k+1]\) for each \(v\in L\) and \(m_v^R: 2^L \to [4]\) for each \(v\in R\), and an algorithm \(\mathcal{A}\) such that, given \((v,m_v(N^{\omega=1}(v)))\) for each vertex \(v\), \(\mathcal{A}\) computes the connected components of \((L\uplus R,E^{\omega=1})\), where \(m_v\) denotes the appropriate map for \(v\in L\uplus R\) among \(m_v^L\) and \(m_v^R\).
\end{lemma}

Using this protocol, vertices on the left will have to transmit \(\lceil\lg{(k+1)}\rceil\) bits, while vertices on the right will only have to transmit \(2\) bits. Thus, the vertices on the right will save a bit each compared to reporting all \(3\) incident edges. This is what leads to the improvement in our protocol.

The second ingredient is a balanced decomposition of a complete graph into such bipartite subgraphs.

\begin{lemma}[store=lem:cover]
    Let $s,t$ be positive integers, and let $G=(V,E)$ be the complete graph
    on $2st+1$ vertices. Then $E$ admits a partition into copies of
    $K_{s,t}$, say with parts $(L_i,R_i)$, such that every vertex occurs
    in exactly $s$ left parts and exactly $t$ right parts.
\end{lemma}

Taking \(s=3\) and \(t=k\) means that we are able to apply \cref{lem:algorithm}, thus giving that each vertex occurs \(3\) times on the left, which contributes \(\lceil 3\lg{(k+1)}\rceil\) bits, while occurring \(k\) times on the right, contributing \(2k\) bits. The graph we apply this theorem to will have \(6k+1\) vertices, which \(2k\) is roughly \(1/3\) of.

\begin{theorem}
    Given a graph \(G=(V,E)\) with \(|V|=n\), there exists a collection of maps \(r_v: 2^V\to \{0,1\}^t\) for \(n \ge 13\) with \(t=\frac{n}{3} + 3\lg n\) such that given \(\{(v,r_v(N(v))\}_{v\in V}\) one can uniquely determine the connected components of \(G\).
\end{theorem}
\begin{proof}
    Let \(k=\ceil{\frac{n-1}{6}}\) and \(n'=6k+1\).
    First, apply \cref{lem:cover} on \(K_{n'}\) with \(s=3\) and \(t=k\). This gives a partition of the edges into \((L_i\uplus R_i, E_i)_i\) with \(|L_i|=3\) and \(|R_i|=k\), such that each vertex occurs \(3\) times as a vertex in some \(L_i\) and \(k\) times as a vertex in some \(R_i\).  Now for each vertex \(v\in V\), we identify it with some vertex from \(K_{n'}\). Then, let \(\omega(e) = [e\in E]\) and apply \cref{lem:algorithm} on each of the edge partitions. Finally, let \(r_v\) for some vertex \(v\) be the combined output of each \(m_v\) from the partitions of \((L_i\uplus R_i, E_i)\). Since \(v\) participates three times in \(L_i\)'s and \(k\) times in \(R_i\)'s, we observe by \cref{lem:algorithm} that the total length of the concatenation of messages is 
    \[
        \lceil 3\lg (k+1)\rceil + 2k \le \frac{n}{3} + 3\lg n
    \]
    for \(n \ge 13\).  One thing to note in this calculation is that to get the \(3\) on the left side inside the ceiling function, we encode the degrees together. There are then \((k+1)^3\) different possible degree triples, which requires \(3\lg (k+1)\) bits to encode, resulting in the presented expression. For all vertices of \(K_{n'}\) that were not identified with any vertices of \(G\), we know all of them are singletons, and so their messages can be easily simulated by the algorithm. Finally, applying \cref{lem:uncover} gives the connected components of \(G\).
\end{proof}

The theorem is primarily of interest because it shows that for computing either connected components or connectivity of a graph, it is possible to have a protocol that performs better than the naive one, which was previously suspected could be optimal. The authors of \cite{hktzz19} ask if the naive protocol for producing spanning forest is optimal, and at least for producing connected components, this shows it is not.

As an interesting note, while previous protocols generally have solved spanning forest which gives connected components and connectivity \cite{agm12linear,agm12sketch,hktzz19}, we only solve the problem of connected components. In fact, it is in general not possible to extract a spanning forest from the messages that are transmitted in our protocol.

Finally, we also show that the question of determining the connected components in a graph indeed is equivalent up to a factor 2 in the size of the graph, with a polynomial blow-up in the error rate: 
\begin{theorem}[store=thm:equivalence]
    Let $\mathcal{S}$ be a randomized neighbourhood-sketching scheme for
    connectivity. Suppose that, on every graph of $N$ vertices, each
    vertex sends a sketch of at most $t(N)$ bits and the referee decides
    connectivity with error probability at most $\delta(N)$.

    Then there is a randomized neighbourhood-sketching scheme that, on every
    $n$-vertex graph, outputs its connected components using sketches of
    length at most $t(2n)$ and with error probability at most
    \[
        (n-1)^2\delta(2n).
    \]
    In particular, connectivity sketches with error probability of at most
    $N^{-(c+2)}$ yield connected-component sketches with error probability of at most
    $n^{-c}$.
\end{theorem}
Another way to frame this result is that if we let \(C_{cc}(n)\) denote the minimum message length required per vertex to compute the connected components of graphs with \(n\) vertices and let \(C_{conn}(n)\) be the minimum message length required to compute whether the graph is connected, then \(C_{cc}(n)\le C_{conn}(2n)\) with a multiplicative blow-up in the error probability of \((n-1)^2\). 

The proof of this statement is shown by having the vertices of the graph compose messages as if the graph has double size, and each vertex is connected to some mirror vertex, that doesn't exist in the original graph. The referee can then connect the mirror vertices as they want and simulate the resulting messages, and then query the connectivity algorithm on these messages.

\subsection{Related Work}

\paragraph{Public Randomness} As mentioned previously, graph sketching has been studied in various settings related to randomness. In the setting where all vertices have access to shared randomness, the framework by Ahn, Guha, and McGregor \cite{agm12linear,agm12sketch} allow each vertex to send messages of length at most \(O(\log^3 n)\) bits per vertex for the purpose of computing a spanning forest. Their sketch however only works in the case that the vertices need to have access to a shared source of randomness. This is due to the fact that they employcoordinated sampling of edges, which need to match between vertices. 

Later Nelson and Yu \cite{nelson2019lowertree} showed that this was tight for computing spanning forests, requiring \(\Omega(\log^3 n)\) bits on average per vertex. This was subsequently strengthened by Yu \cite{yu2021connectivity} to show that this lower bound even holds for the problem of determining if the graph is connected. Both of these results rely only on public randomness, and as such it remains open if restricting to private randomness or deterministic protocols could strengthen the lower bounds.

\paragraph{Private Randomness} The reliance on public randomness was subsequently removed in exchange for significantly larger message sizes by Holm, King, Thorup, Zamir, and Zwick \cite{hktzz19}. They showed that by having each vertex sample \(O(\sqrt{n})\) neighbours independently of each other, and report these, only \(O(\sqrt{n})\) edges remain between the connected components induced by the reported edges. The remaining edges could then be determined using error-correcting coding related techniques. 

\paragraph{Deterministic protocols and lower bounds.}
The straightforward deterministic protocol fixes a balanced orientation of \(K_n\) and has each vertex report the presence or absence of the potential edges assigned to it, using at most \(\lfloor n/2\rfloor\) bits per vertex.  The authors of \cite{hktzz19}noted that no better deterministic protocol was known for producing a spanning forest and asked whether this bound is optimal.  Our result beats this bound for the weaker task of recovering the connected-component partition, but it does not identify a spanning forest.

Robinson and Tan recently obtained the first super-polylogarithmic deterministic lower bound for a connectivity decision problem: deciding \(k\)-edge connectivity requires worst-case messages of \(\Omega(k)\) bits for every super-constant \(k=O(\sqrt n)\)~\cite{robinson2026detlowerkcon}.  This result does not yield a polynomial lower bound for the ordinary \(k=1\) case.

For ordinary connectivity, Yu's \(\Omega(\log^3 n)\) public-coin lower bound also holds for deterministic protocols~\cite{yu2021connectivity}. That lower bound has, in the deterministic setting, very recently been improved by Robinson and Tan \cite{robinson2026polynomiallowerboundsdistributed}, who showed a polynomial lower bound of $\Omega(n^{1/3})$ bits per vertex.

Finally, a similar reduction between connected components and connectivity as ours has previously been shown by Lizama and Rapaport \cite{Lizama2022}. Using a construction with a number of similarities with our reduction, they showed that if \(C_{\# cc}(n)\) is the minimum message length required to compute the number of connected components, then \(C_{\#cc}(n)\le 2C_{conn}(n+2)\). Their bound, however only holds for protocols that are either deterministic or have exponentially small error probability.

\section{Bipartite graph}
In the following lemma, for the set \(S\) of edges let $S^{\omega=1}$ denote the edges labelled with $1$ by \(\omega\). If instead we are considering the set of neighbours \(N(v)\) of some vertex \(v\), we let \(N^{\omega=1}(v)\) be the set \(\{w\in N(v) \; | \; \omega(v,w)=1\}\).
\getkeytheorem{lem:algorithm}
\begin{proof}
Let \(L=\{l_0,l_1,l_2\}\), with indices taken modulo \(3\). For \(l_i\in L\), let
\[
    m_{l_i}(N^{\omega=1}(l_i))=|N^{\omega=1}(l_i)|.
\]
This requires \(k+1\) possible values. For \(r\in R\), put \(S_r=N^{\omega=1}(r)\subseteq L\). If \(S_r=L\), output \(0\).  Otherwise, if there is an \(i\in\{0,1,2\}\) such that
\[
    l_i\in S_r
    \qquad\text{and}\qquad
    l_{i+1}\notin S_r,
\]
output such an \(i\). Finally, if \(S_r=\emptyset\), output \(3\). Thus, each right vertex sends one of four possible messages.

We first describe the reconstruction under the simplifying assumption that no right vertex has degree \(3\), i.e. \(S_r\neq L\) for all \(r\in R\). Let
\[
    d_i:=|N^{\omega=1}(l_i)|
\]
be the message of \(l_i\), and let
\[
    n_i:=|\{r\in R:m_r(S_r)=i\}|
\]
for \(i=0,1,2\) be the number of vertices reporting message \(i\). Vertices with message \(3\) are isolated.

Under the assumption \(S_r\neq L\), the messages have the following meaning:
\[
\begin{array}{c|c}
\text{message} & \text{possible neighbourhoods} \\ \hline
0 & \{l_0\},\{l_0,l_2\} \\
1 & \{l_1\},\{l_0,l_1\} \\
2 & \{l_2\},\{l_1,l_2\} \\
3 & \emptyset .
\end{array}
\]
Hence \(l_0\) has a common neighbour with \(l_1\) iff some message-\(1\) vertex is also adjacent to \(l_0\), which occurs iff
\[
    d_0-n_0>0.
\]
Similarly, \(l_1\)  has a common neighbour with \(l_2\) iff
\[
    d_1-n_1>0,
\]
and \(l_0\)  has a common neighbour with \(l_2\) iff
\[
    d_2-n_2>0.
\]
Thus the messages determine the graph on \(L\) whose edges record which pairs of vertices in \(L\) have a common neighbour in \(R\). The connected components of this auxiliary graph are exactly the connected components of \((L\uplus R,E^{\omega=1})\) restricted to \(L\).

Finally, each \(r\in R\) with message \(m_r\in\{0,1,2\}\) is adjacent to \(l_{m_r}\), so \(r\) belongs to the component of \(l_{m_r}\). Vertices with message \(3\) are isolated. Hence, the components are determined.

It remains only to drop assumption of there being no degree $3$ right vertices. We will use the exact same decoding process, and argue that it is still correct. If some right vertex has degree \(3\), then it is adjacent to all of \(L\), so all three vertices of \(L\) lie in the same connected component. Such vertices are encoded by message \(0\), which is harmless: the above reconstruction may only underestimate pairwise connections inside \(L\), but the presence of a degree-\(3\) vertex forces all of \(L\) into one component. If we consider the previous analysis however, we see that any degree \(3\) vertex implies that $d_1-n_1>0$ and $d_2-n_2>0$, and so all vertices from $L$ are determined to belong to the same connected components.

Consequently, the collection of messages uniquely determines the connected components of \((L\uplus R,E^{\omega=1})\).
\end{proof}

\section{Covering a complete graph}
In this section, we are going to show how to partition the edges of our graph into the bipartite subgraphs required by \cref{lem:algorithm}. We will show a simple explicit way to do this with the following lemma.
\getkeytheorem{lem:cover}

\begin{proof}
    Let $n=2st+1$ and identify $V$ with $\mathbb Z_n$. Define
    \[
        A=\{t,2t,\ldots,st\},\qquad
        B=\{0,1,\ldots,t-1\}.
    \]
    Then $|A|=s$, $|B|=t$, and $A\cap B=\varnothing$. Moreover, the
    differences $a-b$, with $a\in A$ and $b\in B$, are precisely
    $1,2,\ldots,st$ with each value occuring once. For each $x\in\mathbb Z_n$, let
    \[
        L_x=x+A,\qquad R_x=x+B,
    \]
    and let $E_x$ be the edge set of the complete bipartite graph between
    $L_x$ and $R_x$. We claim that the sets $E_x$, $x\in\mathbb Z_n$,
    partition the edges of $K_n$.

    Given an edge $\{u,v\}$, orient it so that
    \[
        v-u\equiv d\pmod n
    \]
    with $d\in\{1,\ldots,st\}$. By the preceding paragraph there is a
    unique pair $(a,b)\in A\times B$ with $a-b=d$. Then with
    $x=u-b$ we have
    \[
        \{u,v\}=\{x+b,x+a\}\in E_x.
    \]
    The same uniqueness of $(a,b)$ also shows that this representation is
    unique, and hence the $E_x$ are pairwise disjoint and cover all edges.

    To show the correct number of occurrences in each left and right part, a fixed vertex $v$ lies in $L_x$ exactly for the $s$ values
    $x=v-a$ with $a\in A$, and lies in $R_x$ exactly for the $t$ values
    $x=v-b$ with $b\in B$. Thus the required incidence condition holds.
\end{proof}

\section{Equivalence between connected components and connectivity}
Finally, we will show that computing connectivity is essentially as hard as computing connected components. This is mostly relevant from the perspective of lower bounds, since it means that it is enough for most use-cases to only consider lower bounds for connected components, and then lower bounds for connectivity will immediately follow.
\getkeytheorem{thm:equivalence}

\begin{proof}
    For a labelled vertex set \(U\), write
    \(m_{U,v}(S)\) for the random sketch produced by vertex \(v\) when
    its neighbourhood is \(S\subseteq U\setminus\{v\}\). Let \(G=(V,E)\), let \(V'=\{v':v\in V\}\) be a disjoint copy of \(V\), and set
    \(\widehat V=V\cup V'\). Each vertex \(v\in V\) sends
    \begin{equation} \label{eq:reduction-message}
        m_{\widehat V,v}\bigl(N_G(v)\cup\{v'\}\bigr).        
    \end{equation}

    This sketch has length at most \(t(2n)\). For any \(F\subseteq\binom{V}{2}\), let
    \[
        F'=\bigl\{\{u',v'\}:\{u,v\}\in F\bigr\}
    \]
    and define
    \[
        H_F=
        \left(
            \widehat V,\,
            E\cup\{\{v,v'\}:v\in V\}\cup F'
        \right).
    \]
    The neighbourhood in \(H_F\) of every original vertex \(v\in V\) is exactly
    the one used in \cref{eq:reduction-message}, independently of \(F\). Moreover, the referee knows
    the neighbourhood of every copied vertex:
    \[
        N_{H_F}(v')
        =
        \{v\}\cup\{u':\{u,v\}\in F\}.
    \]
    It can therefore generate the sketches of all vertices in \(V'\) and run
    the connectivity decoder for \(\mathcal{S}\) on \(H_F\).

    Contracting the edges \(\{v,v'\}\) gives
    \[
        H_F\text{ is connected}
        \quad\Longleftrightarrow\quad
        (V,E\cup F)\text{ is connected}.
    \]
    Thus the received sketches give the referee a randomized connectivity
    oracle for every graph \((V,E\cup F)\), with error at most \(\delta(2n)\)
    per query.

    Fix any spanning tree \(T\) on the labelled set \(V\). Starting with
    \(F=E(T)\), process its edges one at a time, deleting \(e\) whenever the
    oracle says that \((V,E\cup(F\setminus\{e\}))\) remains connected. If all
    answers are correct, the resulting \(F\) is inclusion-minimal subject to
    \((V,E\cup F)\) being connected. After contracting each connected
    component of \(G\), the edges of \(F\) therefore form a tree. In particular,
    \[
        |F|\leq n-1.
    \]

    It remains to identify the cut associated with each \(e=\{a,b\}\in F\).
    Removing \(e\) from \((V,E\cup F)\) produces exactly two connected
    components, say \(A_e\) and \(B_e\), with \(a\in A_e\) and \(b\in B_e\). For
    every \(x\in V\setminus\{a,b\}\), query whether
    \[
        \bigl(V,E\cup(F\setminus\{e\})\cup\{\{a,x\}\}\bigr)
    \]
    is connected. The answer is yes exactly when \(x\in B_e\), so these
    queries recover the cut \((A_e,B_e)\).

    Assign each vertex the signature
    \[
        \sigma(x)=\bigl(\mathbf 1[x\in B_e]\bigr)_{e\in F}.
    \]
    Two vertices have the same signature if and only if they belong to the
    same connected component of \(G\): vertices in one component cannot be
    separated by deleting an edge of \(F\), while any two distinct contracted
    components are separated by some edge of the tree formed by \(F\).
    Hence, the signatures determine the connected-component partition.

    The reverse-deletion step uses \(n-1\) oracle queries, and recovering the
    cuts uses at most \((n-1)(n-2)\) more. Thus an error-free execution makes
    at most
    \[
        (n-1)+(n-1)(n-2)=(n-1)^2
    \]
    queries. Although the queries are adaptive, before the first incorrect answer they coincide with the fixed error-free execution. A union bound over the possible first erroneous query therefore gives failure
    probability at most
    \[
        (n-1)^2\delta(2n).
    \]

    Finally, if \(\delta(N)\leq N^{-(c+2)}\), then
    \[
        (n-1)^2\delta(2n)
        \leq n^2(2n)^{-(c+2)}
        \leq n^{-c}.
    \]
\end{proof}

\section*{Acknowledgements}
Both authors were supported by VILLUM Foundation Grant 54451, Basic Algorithms Research Copenhagen (BARC).

\noindent\begin{minipage}[t]{\dimexpr\linewidth-5.5cm\relax}%
\vspace{0pt}%
\indent D.R.L was supported by the European Research Council (ERC) under the European Union's Horizon 2020 research and innovation programme (Grant agreement No.\ 101019564).
\end{minipage}
\begin{minipage}[t]{5cm}%
\vspace{0pt}%
\includegraphics[width=5cm]{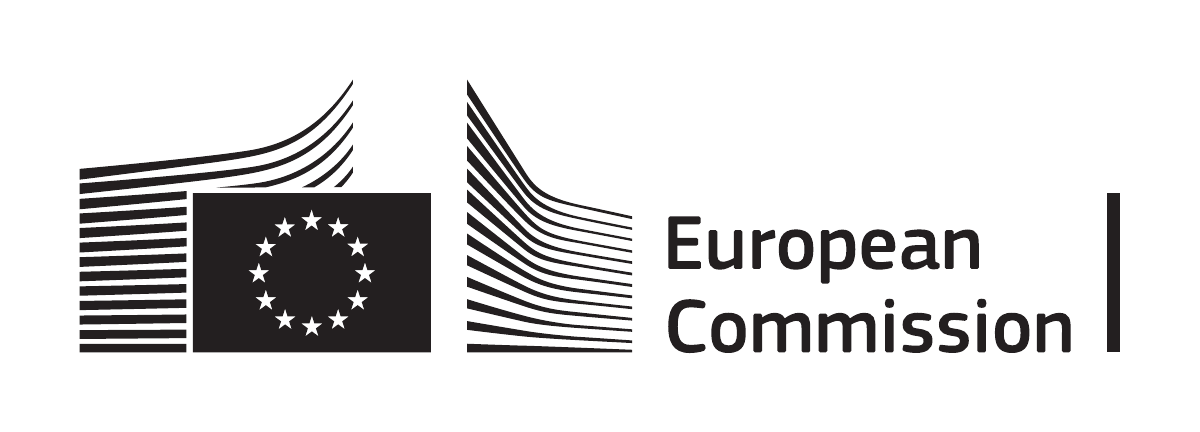}%
\end{minipage}
\smallskip

\noindent For open access purposes, the authors have applied a CC BY public copyright license to any author-accepted manuscript version arising from this submission. Views and opinions expressed are however those of the author(s) only and do not necessarily reflect those of the European Union or the European Research Council Executive Agency. Neither the European Union nor the granting authority can be held responsible for them.

\bibliographystyle{alphaurl}
\bibliography{reference}

\end{document}